# The critical current of YBa$_2$Cu$_3$O$_{7-\delta}$ Low-Angle Grain Boundaries

J. H. Durrell[a], M. J. Hogg, F. Kahlmann, Z. H. Barber, M. G. Blamire, and J. E. Evetts

University of Cambridge, Department of Materials Science and Metallurgy, Pembroke Street, CAMBRIDGE, CB2 3QZ, UK.

Transport critical current measurements have been performed on 5º [001]-tilt thin film YBa$_2$Cu$_3$O$_{7-\delta}$ single grain boundaries with magnetic field rotated in the plane of the film, $\phi$. The variation of the critical current has been determined as a function of the angle between the magnetic field and the grain boundary plane. In applied fields above 1 T the critical current, $j_c$, is found to be strongly suppressed only when the magnetic field is within an angle $\phi_k$ of the grain boundary. Outside this angular range the behavior of the artificial grain boundary is dominated by the critical current of the grains. We show that the $\phi$ dependence of $j_c$ in the suppressed region is well described by a flux cutting model.

74.60.Jg, 74.72.Bk

[a] Corresponding author; email: jhd25@cam.ac.uk

Grain boundaries (GB) in the high-$T_c$ superconductor YBa$_2$Cu$_3$O$_{7-\delta}$ are associated with strongly suppressed values of the transport critical current [1-3]. The reduction in critical current is associated with suppression of the order parameter by strain fields and charging at the grain boundary [4-7]. Two types of grain boundary in YBa$_2$Cu$_3$O$_{7-\delta}$ have been identified. In 'high-angle' grain boundaries the order parameter is continuously suppressed along the boundary; a high-angle GB may therefore be considered as a homogenous barrier and this property has been exploited to produce Josephson junctions. For grain boundary angles <10º however there remains regions where the superconductivity is not suppressed and the order parameter is continuous across the boundary [8, 9]. It is these 'low-angle' grain boundaries which limit the performance of technical conductors.

Modeling of low-angle GB properties suggests that the vortex structure in the grain boundary takes the form of Abrikosov-Josephson (A-J) vortices which are more weakly pinned than vortices in the bulk [10]. Channeling of vortices along the grain boundary has been observed [11,12] as well as pinning of grain boundary vortices by dislocation cores [13]. Although the intergranular critical current, $j_{cGB}$, is reported by many authors [14] to decrease exponentially with increasing grain boundary misorientation angle some authors have suggested a low-angle plateau with $j_{cGB} \sim j_{cIG}$, the in-grain (IG) critical current, for small misorientation angles [15].

Previous studies of low-angle GB have measured the variation of critical current with field applied either solely along the $c$-axis direction or as a function of the angle between the applied field and the $c$-axis, generally with the field parallel to the grain boundary. In the present study we have performed high resolution angular measurements of the variation of the transport critical current with fields rotated about the *c*-axis and aligned with the *a-b* planes.

In this Letter we show that the dependence of the critical current on applied magnetic field for a track containing a low-angle grain boundary is strongly dependent on the angle, $\phi$, between the GB and the magnetic field. We show that this behavior can be described as a cross-over from a high $j_c$ region limited by the pinning within the grains to a region of depressed $j_c$ when the applied field is within a certain angle $\phi_k(B,T)$ of the GB plane.

150 nm thick films of $YBa_2Cu_3O_{7-\delta}$ were grown on 4.9º misorientation [001] tilt $SrTiO_3$ bi-crystal substrates. The films were then patterned to provide 10 µm wide tracks for four point electrical measurements both within the grain (IG tracks) and crossing the grain boundary perpendicularly (GB tracks). Patterning was performed using photolithography and Ar ion-milling. Magnetic field- and angle dependent- critical current measurements were carried out using a two-axis goniometer mounted in an 8 T magnet [16]. The angular accuracy is better than 0.2º, temperature accuracy is better than 20 mK and the field was controlled to better than 0.1 mT. Critical currents were extracted from current voltage (*IV*) characteristics; a voltage criterion of 1 µV was employed, however the form of the observed $j_c(\phi)$ characteristics did not change when the recorded *IV* curves were analyzed using criteria from 0.5 µV to 2.5 µV.

Figure 1 shows the measurement geometry and the results obtained comparing the $j_c(\phi)$ properties of an IG and a GB track. It can be seen that, for larger values of $\phi$, the form of $j_c(\phi)$ is the same in the GB and IG cases. We may therefore define an angular range, $|\phi|<\phi_k$, outside which the angular variation of critical current in the grain boundary track exhibits the same qualitative $j_c(\phi)$ characteristic as the IG track. It can be seen that the presence of the grain boundary only leads to a strongly suppressed critical current for $|\phi|<\phi_k$. In Fig. 1 the IG track $j_c$ for $|\phi|>\phi_k$ is 1.3 times the GB track values. On other samples the grain boundary limited critical current, $j_{cGB}$, has been observed to be equal to the intra-grain critical current, $j_{cIG}$, for $|\phi|>\phi_k$.

Figure 2 shows how the $j_c(\phi)$ behavior develops for varying fields and temperatures. It is clear from Fig 2. that the angular extent of the region of strong $j_c$ suppression increases with decreasing field and temperature.

Grain boundary current transport is associated with a change in the form of the measured voltage versus current characteristics. In previous studies [10, 12] it has been demonstrated that with the field applied parallel to the grain boundary only a few rows of vortices are depinned. This leads to a linear *IV* transition. For the sample studied this behavior is also observed, but only for $|\phi|<\phi_k$. This is shown in Fig. 3. The increased noise in the linear transition is characteristic of vortex flow at a grain boundary as has previously been observed [11]. It is normally assumed in measurements on $YBa_2Cu_3O_{7-\delta}$ grain boundaries that $j_{cGB} << j_{cIG}$. However the results presented here show that behavior typical of a grain boundary is only observed when the magnetic field is applied with $|\phi|<\phi_k$. For $|\phi|>\phi_k$ the critical current behavior of the grain boundary is dominated by the pinning of flux lines within the grains. Only for $|\phi|<\phi_k$ does the reduced pinning of the grain boundary affect the observed critical current behavior of the system and the system then probes the critical current at the grain boundary. Physically therefore $\phi_k$ represents the point for a GB track at which the critical current supportable by the grain boundary $j_{cGB}$ becomes less than the critical current supportable in-grain, $j_{cIG}$. The data in the $j_c(\phi)$ plot shown in Fig. 1 for a GB track therefore represents the lesser of $j_{cIG}(\phi)$ and $j_{cGB}(\phi)$ for any particular value of $\phi$. It should be noted that the results from the IG tracks measure $j_{cIG}(\phi)$.

We may therefore identify three distinct mechanisms which determine $j_c$ depending on the value of $\phi$. For the particular case of $\phi=0°$ and assuming a straight GB the flux is channeling in the region of weak pinning at the grain boundary and the pinning force density on the *A-J* vortices in the GB is directly probed [13]. In practice GB meander leads to a certain amount of extra pinning. For $|\phi|>0°$ the vortices are pinned strongly in the grains and weakly within the GB. However such a GB vortex segment is able to support a larger transport current than predicted from the weak GB pinning. Attachment to the strongly pinned IG parts of the vortex allows the segment to distort without exceeding the critical state and thus carry a larger stable transport current [17]. This additional configurational transport current has a maximum value determined by the onset of phase slip or flux cutting in the vortex system. There are two quite different critical state scenarios. For $|\phi|>\phi_k$ there are a sufficient number of GB vortex segments spanning the GB to transport additional current up to at least $j_{cIG}$. In this case the critical current is determined by the onset of IG flux motion. However for $|\phi|<\phi_k$ the GB vortex segments spanning the GB are fewer in number and each GB segment is unable to support sufficient configurational transport current to allow the IG critical state to be reached. In this case the critical current is limited by $j_{cGB}$, which includes the weak GB flux pinning and also a configurational component arising from the flux vortex distortion. In this case flux flow is localized at the GB and necessarily involves phase slip or flux cutting and cross joining.

When $|\phi|<\phi_k$ the limiting force density on GB vortex segments at the onset of flux cutting may be incorporated into the Lorentz force balance equation:

$$\mathbf{j}_c \times \mathbf{B} = \mathbf{F}_{pin} + \mathbf{F}_{cut} \quad (1)$$

where $\mathbf{F}_{pin}$ is the pinning force density on the segments of vortex within the GB and $\mathbf{F}_{cut}$ is a summation of the flux cutting force over all vortices intersecting the GB expressed as a force density.

In a thin film with strong pinning we may assume that the force on the lattice is simply the sum of the individual forces on vortex lines. We assume that the vortex segments in the grain boundary lie in the same direction as the vortices in the grain, this second assumption is necessary to avoid an unphysically high local density of vortex segments in the grain boundary.

If $L$ is the length of each vortex segment in the grain boundary, given by $L=d_{gb}/\sin|\phi|$ where $d_{gb}$ is the width of the grain boundary, $\mathbf{F}_{pin}$ may be written in terms of the pinning force per unit length on each flux line, $f_{pin}$, and the vortex spacing, $a_0$ [18]:

$$\left|\mathbf{F}_{pin}\right| = \frac{f_{pin}}{a_0^2} \quad (2)$$

Similarly $\mathbf{F}_{cut}$ maybe written in terms of the force required to cut a single grain boundary vortex $f_{cut}$. The total contribution to $\mathbf{F}_{cut}$ will depend on the density of points where vortices cross into the grain boundary, which is $\sin|\phi|/d_{gb}a_0^2$, leading to the following equation:

$$\left|\mathbf{F}_{cut}\right| = \frac{1}{d_{gb}a_o^2} \sin|\phi| f_{cut} \quad (3)$$

The value of $f_{cut}$ may depend on temperature but not on field if an analysis which treats individual vortex lines separately is valid.

Since the GB is perpendicular to the current we may therefore rewrite Eq. 1 as:

$$j_c = f_{pin} \frac{1}{\Phi_0 \cos\phi} + f_{cut} \frac{1}{d_{gb}\Phi_0} \tan|\phi| \qquad (4)$$

The first term of Eq. 4 represents the contribution to the critical current from pinning in the GB and the second the extra pinning due to the fact that the vortex segment in the grain boundary is strongly pinned at either end. It can be seen that at $\phi=0°$ we recover $f_{pin}$, although in practice the fact that the grain boundary meanders means that $f_{pin}$ will be overestimated.

The field dependence of $j_c(\phi)$ for $|\phi|<\phi_k$ is found to be quite different to the temperature dependence. The insets to Fig. 2a and 2b show $j_c(\phi)-(j_c(\phi=0)/\cos|\phi|)$ versus $\phi$; from Eq. 4 this is the critical current component due to flux cutting, $j_{c,cut}$. It can be noted that over a wide range of field (1-8 T) the form of $j_{c,cut}$ remains the same and it is only the point at which $j_{cIG}$ becomes lower than $j_{cGB}$ that determines $\phi_k$.

In contrast the form of $j_{c,cut}$ shows a strong variation with temperature indicating that the enhanced pinning of the vortex lines depends on temperature. This is consistent with a change with varying temperature of the core size of both the IG Abrikosov vortices and the GB *A-J* vortices. This changes the properties of the vortex line thus changing the cutting force.

The following expression for $f_{cut}$ may be obtained from Eq. 4:

$$f_{cut} = \frac{d_{gb}\Phi_0}{\cos\phi} \cdot \frac{d}{d\phi}(j_c \cos\phi) \qquad (5)$$

This yields values for the vortex cutting force from the experimentally measured parameter, $j_c$, without having to correct for the low angle meandering of the GB. We take here a typical value of $d_{gb}$ as being ~10nm. Fig. 4 shows how this force varies with $\phi$. It is as expected, essentially constant outside the region $|\phi|<6º$ where the meander of the grain boundary is significant [19]. The above analysis is only valid for $|\phi|<\phi_k$ since for $|\phi|>\phi_k$ the critical current is determined by IG vortex motion.

In conclusion, we have measured the critical current properties of $YBa_2Cu_3O_{7-\delta}$ low-angle grain boundaries where the applied field is rotated in the plane of the film. We make the striking observation that the critical current is strongly suppressed only when the applied magnetic field is within an angle $\phi_k$ of the grain boundary. However, for small fields (<1 T) the range of $\phi_k$ for which GB behavior is seen is large and the grain boundaries dominate the critical current behavior. These results may well explain reports of $j_{cGB}=j_{cIG}$ in the literature and the low angle plateau sometimes reported in characterizations of the $j_c$ versus misorientation angle properties of $YBa_2Cu_3O_{7-\delta}$ grain boundaries. We have furthermore elucidated the temperature and field dependence of $j_c(\phi)$ in the GB dominated region and found that the extra cutting force depends linearly on temperature and only weakly on field. Implicit in these results is the observation that optimizing the in-grain properties of $YBa_2Cu_3O_{7-\delta}$ coated conductors will affect the observed critical current.

The authors wish to acknowledge financial support from the Engineering and Physical Sciences Research Council.

To be published in PRL

References

[1] D. Dimos, P. Chaudhari, J. Mannhart, and F. K. LeGoues, Phys. Rev. Lett. **61**, 219 (1988).

[2] P. Chaudhari, et al., Phys. Rev. Lett. **60**, 1653 (1988).

[3] D. Dimos, P. Chaudhari, and J. Mannhart, Phys. Rev. B **41**, 4038 (1990).

[4] M. F. Chisholm and S. J. Pennycook, Nature **351**, 47 (1991).

[5] A. Gurevich and E. A. Pashitskii, Phys. Rev. B **57,** 13878 (1998).

[6] H. Hilgenkamp and J. Mannhart, Rev. Mod. Phys. **74**, 485 (2002).

[7] J. Mannhart and H. Hilgenkamp, Mater. Sci. Eng. B **56**, 77 (1998).

[8] D. M. Feldmann, et al., Appl. Phys. Lett. **79**, 3998 (2001).

[9] R. D. Redwing, et al., Appl. Phys. Lett. **75**, 3171 (1999).

[10] A. Gurevich, et al., Phys. Rev. Lett. **88**, 097001 (2002).

[11] M. J. Hogg, F. Kahlmann, E. J. Tarte, Z. H. Barber, and J. E. Evetts, Appl. Phys. Lett. **78**, 1433 (2001).

[12] A. Diaz, L. Mechin, P. Berghuis, and J. E. Evetts, Phys. rev. B **58**, R2960 (1998).

[13] A. Diaz, L. Mechin, P. Berghuis, and J. E. Evetts, Phys. Rev. Lett. **80**, 3855 (1998).

[14] D. Larbalestier, et al., Nature **414**, 368 (2001).

[15] N. F. Heinig, R. D. Redwing, J. E. Nordman and D. C. Larbalestier, Phys. Rev. B **60**, 1409 (1999).

[16] R. Herzog and J. E. Evetts, Rev. Sci. Instrum. **65**, 3574 (1994).

[17] In a large system of vortices current flow between separated strongly pinned regions is achieved by allowing vortices to take a force free configuration described by

$j_c \times B=0$. In a small system this is conveniently viewed as current transport across a region of low or zero pinning with a corresponding elastic bowing of the free vortex segments stabilised by strong pinning at the boundary of the region.

[18] The lattice is disordered so the definition of $a_0$ here is the average spacing between vortices. We therefore assume that $B=\Phi_0/a_0^2$.

[19] The grain boundary was examined with atomic force microscopy and as a first approximation the scatter of orientation of facets along the grain boundary was estimated to be ~ 6 degrees.

To be published in PRL

Figure Captions

**Fig. 1. Critical current dependence on magnetic field angle, $\phi$, for an in-grain (IG) track and a grain boundary crossing track (GB). The inset defines the field rotation angle, $\phi$, and indicates the direction of the Lorentz force, $F_L$, on the flux lines. The angle at which the behaviour of the GB track deviates from the in-grain behaviour is defined as $\phi_k$.**

**Fig. 2. GB track critical current versus magnetic field orientation shown for varying fields (a) and temperatures (b). The insets show $j_{c,cut}$, the critical current with the contribution due to pinning of flux in the grain boundary subtracted for $\phi<\phi_k$.**

**Fig. 3 Critical current transitions recorded with the field oriented along the grain boundary ($\phi=0$, lower scale) and away from the grain boundary ($\phi=-40$, upper scale). The inset shows how the dV/dI characteristic, measured at the criterion voltage, changes with $\phi$. It can be seen that the $VJ$ characteristic is much sharper when $\phi<\phi_k$.**

**Fig. 4. Variation of the flux cutting force with $\phi$ for varying temperatures. The inset shows the averaged variation of $f_{cut}$ with temperature.**

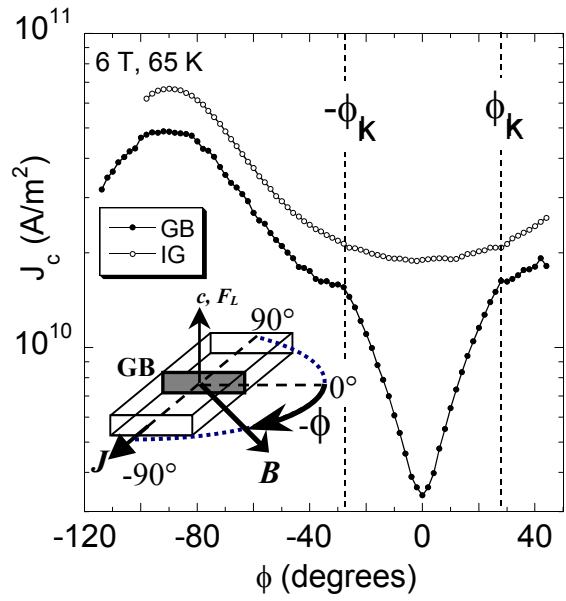

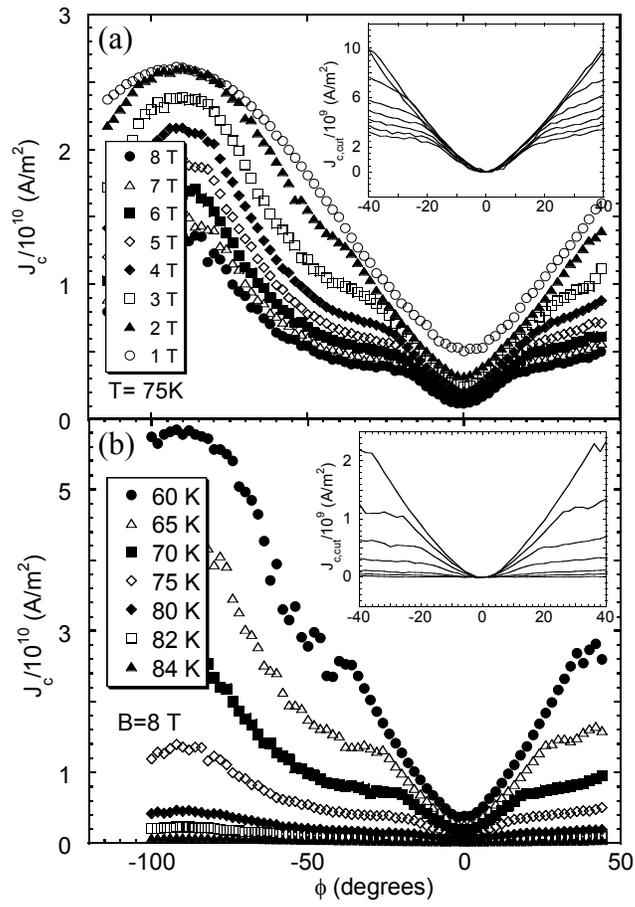

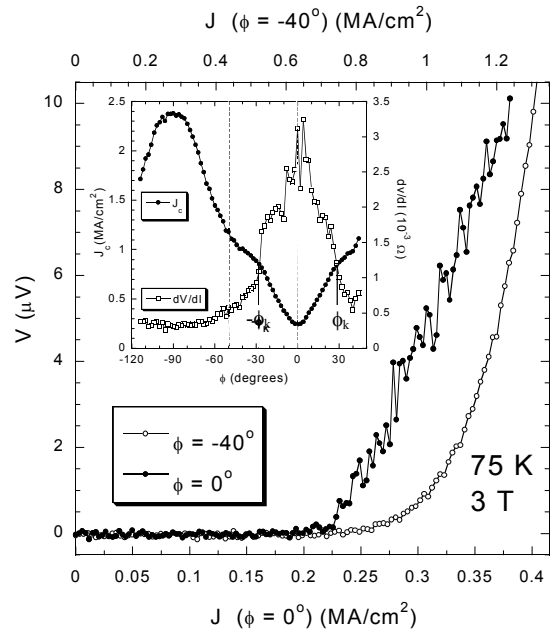

To be published in PRL

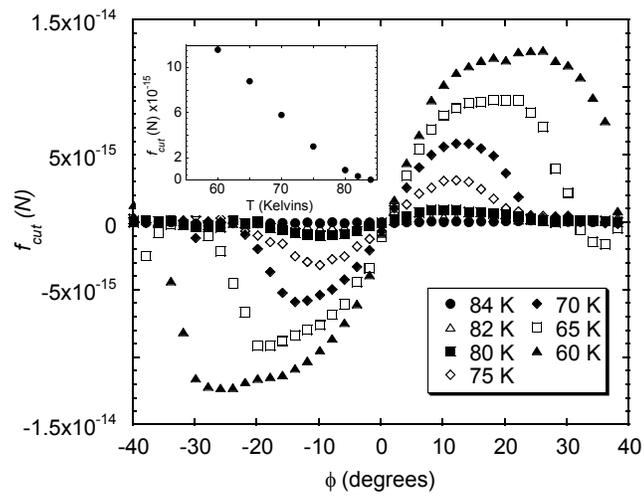